\documentclass[conference]{IEEEtran}

\usepackage{epstopdf}
\usepackage[utf8]{inputenc}

\usepackage{graphicx}
\usepackage{subfig}

\usepackage{textcomp}
\usepackage{enumitem}
\usepackage[font=footnotesize]{caption}

\usepackage{multirow}
\usepackage{subcaption}
\usepackage{siunitx}
\usepackage{comment}
\usepackage{tabularx}
\usepackage{booktabs}
\usepackage[export]{adjustbox}

\usepackage[table]{xcolor}
\usepackage{soul}
\definecolor{softgray}{gray}{0.96}

\usepackage{cite}
\usepackage{amsmath}
\usepackage{amssymb}
\usepackage{multicol}
\usepackage{mathtools}
\usepackage{etoolbox}

\newcounter{groupeq}
{%
  \refstepcounter{equation}
  \protected@edef\thegroupeqname{\theequation}
  \setcounter{groupeq}{0}%
  \def\theequation{\thegroupeqname.\alph{groupeq}}%
  \def\thegroupequationsname{#1}%
}%
{%
  \def\theequation{\thegroupeqname}%
}

\makeatletter
\@addtoreset{groupeq}{equation}
\makeatother

\usepackage{algorithm}
\usepackage{algpseudocode}

\title{Tempered Christoffel-Weighted Polynomial Chaos Expansion for Resilience-Oriented Uncertainty Quantification}

\author{\IEEEauthorblockN{Mahsa Ebadat-Parast, Xiaozhe Wang}
\IEEEauthorblockA{\textit{Department of Electrical and Computer Engineering},
\textit{McGill University}, Montreal, Quebec, Canada\\
Email: mahsa.ebadatparast@mail.mcgill.ca,xiaozhe.wang2@mcgill.ca}
}
\begin{document}
\maketitle
\renewcommand{\thefootnote}{}
\footnotetext{This work was supported by the Natural Sciences and Engineering Research Council (NSERC) Discovery Grant, NSERC RGPIN-2022-03236, CRC-2023-00006, and by the Rubin \& So Foundation Faculty Scholar Award. (Corresponding author: Xiaozhe Wang.)}
\addtocounter{footnote}{0}


\begin{abstract}
Accurate and efficient uncertainty quantification is essential for resilience assessment of modern power systems under high impact and low probability disturbances. Data driven sparse polynomial chaos expansion (DDSPCE) provides a computationally efficient surrogate framework but may suffer from ill conditioned regression and loss of accuracy in the distribution tails that determine system risk. 
This paper studies the impact of regression weighting schemes on the stability and tail accuracy of DD-SPCE surrogates by introducing a tempered Christoffel weighted least squares (T-CWLS) formulation that balances numerical stability and tail fidelity.
The tempering exponent is treated as a hyperparameter whose influence is examined with respect to distributional accuracy compared with Monte Carlo simulations.
Case studies on distribution system load shedding show that the proposed method reduces 95th percentile deviation by 16\%, 5th percentile deviation by 6\%, and improves the regression stability index by over 130\%. 
The results demonstrate that controlling the weighting intensity directly influences both stability index and the accuracy of tail prediction.
\end{abstract}


\section{Introduction}
\label{sec:introduction}

Increasing high-impact, low-probability (HILP) events such as hurricanes, wildfires, and cyberattacks highlight the need for \emph{resilience} in power systems,defined as the ability to anticipate, absorb, adapt, and recover~\cite{panteli2017power,liu2017quantified}.
{To enable resilience-oriented decision-making, researchers have increasingly relied on probabilistic modeling and uncertainty quantification frameworks~\cite{willems2024probabilistic}. Among them, Monte Carlo simulation (MCS) remains the most accurate but suffers from prohibitive computational cost for large-scale systems. 
Uncertainty in power system resilience has also been modeled through scenario-based and Monte Carlo methods~\cite{li2023restoration,huang2024resilience}. 
Robust and hybrid stochastic–robust optimization frameworks have further been applied to ensure feasible operation under extreme conditions~\cite{dehghani2022multi,jokar2025robust}. While these methods capture representative scenarios or bounded uncertainty regions, they often entail heavy computational cost and may become overly conservative for large-scale systems, motivating the use of surrogate-based uncertainty quantification techniques.}
Among such methods, polynomial chaos expansion (PCE) and its data-driven variants offer an efficient surrogate modeling framework that approximates the stochastic system response using a limited number of deterministic simulations. Data-driven sparse PCE (DD-SPCE) further eliminates the need for explicit probability distributions, enabling surrogate construction directly from available samples or 
{historical data~\cite{wang2023comparative,Wang2021ddspce,wang2021data}.} These features make DD-SPCE particularly attractive for resilience analysis, where data may be limited and system models are computationally intensive to evaluate.

However, the regression system in DD-SPCE can become ill-conditioned when the experimental design is uneven or the sample set is small, resulting in unstable surrogate coefficients and degraded prediction accuracy. 

To improve stability, Christoffel-weighted least squares (CLS) methods have been proposed, assigning inverse leverage-based weights to reduce the influence of high-response samples~\cite{cohen2013stability,narayan2017christoffel}. Although this strategy enhances numerical conditioning, it tends to underestimate output tails. Accurately capturing the distribution tail is essential, as it governs the probability of extreme disruptions.
{To address this trade-off between stability and tail fidelity, this work investigates how different regression weighting schemes within a data-driven sparse PCE framework influence the stability of surrogate predictions. In particular, a tempering factor is introduced to control the weighting intensity between the ordinary least squares (OLS) and the Christoffel-weighted formulations, enabling a systematic study of how weighting strength affects both numerical conditioning and tail representation.}
Using MCS results as a benchmark, the study shows that moderate tempering improves the surrogate’s stability and tail accuracy under limited data. This balance enables more reliable estimation of resilience‐related probabilistic metrics. Overall, adjusting regression weights provides a practical means to control the trade‐off between regression robustness and tail representation in data‐driven polynomial chaos models.

\section{Resilience-Oriented Monte Carlo-Based Optimization Modeling}
\subsection{Model Description and Assumptions}
The system under study comprises three interconnected microgrids (MGs) linked through a common distribution network. Under normal conditions, each MG exchanges active power with the upstream grid via its point of common coupling (PCC). Following an extreme event, the distribution network becomes isolated, and the MGs operate in emergency mode ,representing the system’s resilience response to external disturbances.
Loads in both the distribution system and MGs are categorized into three priority levels,from critical to non-critical,based on their importance and service requirements. During emergencies, MGs should supply critical and high-priority loads, while lower-priority ones may be curtailed if local resources are insufficient.
{
System resilience is quantified by the  total operation cost which includes a penalty term for unserved load in MGs as an objective function and  load shedding in the distribution system as a constraint, a widely adopted metric in resilience studies~\cite{dui2024optimizing,kumari2024multi,wang2018resilience}. Minimizing this cost reflects the system’s capability to maintain essential services during disruptions. Primary uncertainties are considered: (i) load demand, and (ii) emergency duration and (iii) initiation time. A MCS framework generates stochastic realizations of these parameters to capture their probabilistic behavior.}
\subsection{Optimization Formulation}
The  mixed integer linear programming formulation jointly models normal ($\Gamma=0$) and emergency ($\Gamma=1$) operation modes, where $\Gamma$ activates emergency-only components. The objective function in \eqref{eq:1} minimizes the total operating cost, including generation, renewable energy curtailment, storage usage, and market transactions in normal mode, as well as load shedding penalties under emergency conditions. Microgrid-related constraints \eqref{eq:2a}--\eqref{eq:2m} ensure local active power balance, import/export limits, and disconnection logic during isolation; enforce unit commitment feasibility and ramping limits for dispatchable units; and model the charging/discharging dynamics and state-of-charge boundaries of the energy storage systems. Distribution-level equations \eqref{eq:3a}--\eqref{eq:3c} maintain system-wide active and reactive power balance, incorporate the linearized DistFlow voltage relations, and preserve power-factor consistency for curtailed loads. Finally, the coupling constraints \eqref{eq:4a}--\eqref{eq:4c} define the allowable inter-microgrid power exchanges through the PCC links and enforce converter-based coupling limits. Collectively, all constraints and the unified objective yield a comprehensive stochastic scheduling model that captures both normal and emergency operation behavior through the resilience control parameter $\Gamma$.

\begingroup
\setlength{\abovedisplayskip}{2pt}
\setlength{\belowdisplayskip}{2pt}
\setlength{\abovedisplayshortskip}{0pt}
\setlength{\belowdisplayshortskip}{0pt}

\begin{flalign}
& \min \Bigg[
\sum_{t\in T}\sum_{m\in M}\sum_{g\in G_m}
\big(c_g^{op} PG_{m,g,t}^{\omega}
+ c_g^{su} y_{m,g,t}
+ c_g^{sd} z_{m,g,t}\big) && \label{eq:1}\\
&\qquad
+ \sum_{t\in T}\sum_{m\in M}
\big(c_{ren,m} Pren_{m,t}^{\omega}
+ c_{b,m} P_{m,t}^{dis,\omega}\big) && \notag\\
&\qquad
+ (1-\Gamma)\sum_{t\in T}\sum_{m\in M}
\big(\pi_t^{buy} P_{imp,m,t}^{\omega}
- \pi_t^{sell} P_{exp,m,t}^{\omega}\big) && \notag\\
&\qquad
+ \Gamma \sum_{t\in T}\sum_{m\in M}\sum_{k\in K}
\lambda_k^{mg}\, PLsh_{m,t,k}^{\omega}
\Bigg] && \notag
\end{flalign}
\begin{subequations}\label{grp:MG}
\renewcommand{\theequation}{\theparentequation.\alph{equation}}
\begin{flalign}
& 0 \le P_{imp,m,t}^{\omega}
\le \beta_{imp,m,t} P_{imp}^{\max}, 
\quad \forall t \in T_{nr} && \label{eq:2a}
\end{flalign}
\begin{flalign}
& 0 \le P_{exp,m,t}^{\omega}
\le \beta_{exp,m,t} P_{exp}^{\max}, 
\quad \forall t \in T_{nr} && \label{eq:2b}
\end{flalign}
\begin{flalign}
& \beta_{imp,m,t} + \beta_{exp,m,t} \le 1,
\quad \forall t \in T_{nr} && \label{eq:2c}
\end{flalign}
\begin{flalign}
& P_{imp,m,t}^{\omega} = P_{exp,m,t}^{\omega} = 0,
\quad \forall t \in T_{em} && \label{eq:2d}
\end{flalign}
\begin{flalign}
& PG_{m,g,t-1}^{\omega}-PG_{m,g,t}^{\omega}
\le RRD_{g} x_{m,g,t}+SD_{g} z_{m,g,t} && \label{eq:2e}
\end{flalign}
\begin{flalign}
& PG_{m,g,t}^{\omega}-PG_{m,g,t-1}^{\omega}
\le RRU_{g} x_{m,g,t-1}+SU_{g} y_{m,g,t} && \label{eq:2g}
\end{flalign}
\begin{flalign}
& x_{m,g,t} PG_{m,g}^{\min}
\le PG_{m,g,t}^{\omega}
\le x_{m,g,t} PG_{m,g}^{\max} && \label{eq:2h}
\end{flalign}
\begin{flalign}
& y_{m,g,t}+z_{m,g,t}\le 1,\quad
y_{m,g,t}-z_{m,g,t}
= x_{m,g,t}-x_{m,g,t-1} && \label{eq:2i}
\end{flalign}

\begin{flalign}
& SOC_{m,t}^{\omega}
= SOC_{m,t-1}^{\omega}
+ \eta_{m}^{ch} P_{m,t}^{ch,\omega}
- \tfrac{P_{m,t}^{dis,\omega}}{\eta_{m}^{dis}} && \label{eq:2j}
\end{flalign}
\begin{flalign}
& SOC_{m}^{\min} \le SOC_{m,t}^{\omega} \le SOC_{m}^{\max} && \label{eq:2k}
\end{flalign}
\begin{flalign}
& 0 \le P_{m,t}^{ch,\omega} \le I_{m,t}^{ch,\omega} P^{ch,\max},\quad
0 \le P_{m,t}^{dis,\omega} \le I_{m,t}^{dis,\omega} P^{dis,\max} && \label{eq:2l}
\end{flalign}
\begin{flalign}
& I_{m,t}^{ch,\omega} + I_{m,t}^{dis,\omega} \le 1 && \label{eq:2m}
\end{flalign}
\end{subequations}

\begin{subequations}\label{grp:DIS}
\renewcommand{\theequation}{\theparentequation.\alph{equation}}

\begin{flalign}
& V_{n,t}^{\omega}-V_{p,t}^{\omega}
= \frac{R_{n,p} P_{n,p,t}^{\omega}
+ X_{n,p} Q_{n,p,t}^{\omega}}{V_0} && \label{eq:3a}
\end{flalign}
\begin{flalign}
& V_{\min}\le V_{n,t}^{\omega}\le V_{\max},\quad
|P_{n,p,t}^{\omega}|, |Q_{n,p,t}^{\omega}|
\le S_{n,p}^{\max} && \label{eq:3b}
\end{flalign}

\begin{flalign}
& PLsh_{n,t,k}^{dis,\omega}\, Q_{t,n,k}^{dis}
= QLsh_{n,t,k}^{dis,\omega}\, D_{t,n,k}^{dis}, \quad \forall t \in T_{em} && \label{eq:3c}
\end{flalign}
\end{subequations}

\begin{subequations}\label{grp:COUP}
\renewcommand{\theequation}{\theparentequation.\alph{equation}}
\begin{flalign}
& \sum_{g\in G_m} PG_{m,g,t}^{\omega}
+ P_{m,t}^{dis,\omega}
+ Pren_{m,t}^{\omega} && \label{eq:4a}\\
&\qquad 
+ (1-\Gamma)\big(P_{imp,m,t}^{\omega}
- P_{exp,m,t}^{\omega}\big)
+ \Gamma\, P_{ent,m,t}^{\omega} && \notag\\
&\qquad
+ \Gamma \sum_{k\in K} PLsh_{m,t,k}^{\omega}
= \sum_{k\in K} D_{m,t,k}
+ P_{m,t}^{ch,\omega}
+ \Gamma\, P_{md,m,t}^{\omega} && \notag
\end{flalign}
\begin{flalign}
& \sum_{k}\sum_{n} D_{t,n,k}^{dis}
= (1-\Gamma)\!\Big(
P_{sub,t}^{\omega}
+ \sum_{m}\eta_{inv,m} P_{exp,m,t}^{\omega}
- \sum_{m}\tfrac{P_{imp,m,t}^{\omega}}{\eta_{rec,m}}
\Big) && \label{eq:4b}\\
&\qquad
+ \Gamma\!\Big(
\sum_{m}\eta_{inv,m} P_{md,m,t}^{\omega}
+ \sum_{k}\sum_{n} PLsh_{n,t,k}^{dis,\omega}
\Big) && \notag
\end{flalign}

\begin{flalign}
& \sum_{k}\sum_{n} Q_{t,n,k}^{dis}
= \sum_{m} Q_{vsc,m,t}^{\omega}
+ \Gamma \sum_{k}\sum_{n} QLsh_{n,t,k}^{dis,\omega} && \label{eq:4c}
\end{flalign}
\end{subequations}
\endgroup

\section{Proposed Approach}
\label{sec:proposed}
{This section examines how regression weighting schemes within a DDSPCE framework influence numerical conditioning and predictive accuracy. The study compares OLS, Christoffel-weighted least squares, and a \emph{tempered} Christoffel-weighted regression, where a scalar exponent~$\alpha$ controls the weighting intensity. The analysis highlights how the weighting strength affects both surrogate stability and the representation of rare events.
A Monte Carlo dataset is employed as a reference to evaluate tail agreement and to guide the selection of~$\alpha$.}
\subsection{Data-Driven Polynomial Chaos Expansion (DD-PCE)}
\label{sec:ddpce}

Consider $d$ independent input random variables 
$\mathbf{X} = (X_1,\dots,X_d)$ with unknown or partially known
joint probability measure.
Let $\{\mathbf{x}_i\}_{i=1}^{M}$ denote $M$ realizations of $\mathbf{X}$
obtained from simulation or measurement, and let
$\mathbf{Y} = [Y_1,\dots,Y_M]^{\top}$ be the corresponding model evaluations.

\paragraph{Moment-based orthogonal polynomial basis.}
When {analytical probability density functions of the inputs are not available, an orthogonal polynomial basis can be constructed directly from the samples.}
Following the data-driven framework of
Wang~\textit{et al.}~\cite{Wang2021ddspce},
the univariate basis $\{\phi_k(x)\}_{k=0}^{p}$ is obtained by enforcing
the discrete orthogonality condition
\begin{equation}
    \frac{1}{M}\sum_{i=1}^{M} \phi_k(x_i)\,\phi_{\ell}(x_i)
    = \delta_{k\ell},
    \qquad k,\ell=0,\dots,p,
    \label{eq:orth_cond}
\end{equation}
where $\delta_{k\ell}$ is the Kronecker delta.
In practice, the coefficients of $\phi_k(x)$ are determined
by a Stieltjes or Hankel–moment orthogonalization based on the empirical moments
$\mu_r = \tfrac{1}{M}\sum_{i=1}^{M} x_i^r$.

For multivariate inputs, the polynomial basis functions are constructed by
tensorization:
\begin{equation}
    \psi_{\boldsymbol{\nu}}(\mathbf{x})
    = \prod_{j=1}^{d} \phi_{\nu_j}(x_j),
    \qquad
    \boldsymbol{\nu} = (\nu_1,\dots,\nu_d)\in\mathbb{N}_0^d,
    \label{eq:multivariate_basis}
\end{equation}
and a truncation rule (e.g., total degree $p$) limits the number of basis terms
$N=\#\{\boldsymbol{\nu}\}$.

\paragraph{Data-driven PCE representation and OLS formulation.}
The model response is approximated as
\begin{equation}
    Y(\mathbf{x})
    \approx \sum_{j=1}^{N} c_j\,\psi_j(\mathbf{x}),
    \qquad
    \psi_j(\mathbf{x}) \equiv \psi_{\boldsymbol{\nu}_j}(\mathbf{x}),
    \label{eq:pce_expansion}
\end{equation}
where $\boldsymbol{c}=[c_1,\dots,c_N]^{\top}$ are unknown coefficients.
Let the design matrix $\boldsymbol{\Psi}\in\mathbb{R}^{M\times N}$ have entries
$\Psi_{ij}=\psi_j(\mathbf{x}_i)$.  
The coefficients are obtained by solving the
ordinary least-squares  problem
\begin{equation}
    \min_{\boldsymbol{c}}
    \; J_{\mathrm{OLS}}(\boldsymbol{c})
    = \frac{1}{M}\,\|\boldsymbol{\Psi}\boldsymbol{c}-\mathbf{Y}\|_2^2,
    \label{eq:ols_ddpce}
\end{equation}
whose normal equations read
\begin{equation}
    \mathbf{G}\,\boldsymbol{c}
    = \frac{1}{M}\,\boldsymbol{\Psi}^{\top}\mathbf{Y},
    \qquad
    \mathbf{G}
    = \frac{1}{M}\,\boldsymbol{\Psi}^{\top}\boldsymbol{\Psi}.
    \label{eq:ols_normal_eqs}
\end{equation}
Equation~\eqref{eq:ols_ddpce} constitutes the
\emph{ordinary least-squares regression} employed in
the DD-PCE approach of Wang~\textit{et al.}~\cite{wang2021data},
where all samples are equally weighted.
{This unweighted formulation serves as the reference for evaluating how alternative weighting strategies influence the surrogate stability and prediction accuracy in the subsequent analysis.}


\subsection{Christoffel Function and Weighted Least Squares}
\label{sec:cls_theory}

The stability of least-squares polynomial approximation depends on the conditioning
of the empirical Gram matrix~\cite{cohen2013stability}:
\begin{equation}
    \mathbf{G}
    = \frac{1}{M}\boldsymbol{\Psi}^{\top}\boldsymbol{\Psi},
    \label{eq:gram_matrix}
\end{equation}
where $\boldsymbol{\Psi}\in\mathbb{R}^{M\times N}$ is the design matrix with entries
$\Psi_{ij}=\psi_j(\mathbf{x}_i)$.

\vspace{0.2em}
Cohen \textit{et al.}~\cite{cohen2013stability} introduced the \emph{Christoffel function}
to characterize the leverage of each sample point as
\begin{equation}
    K_i
    = \big\|\mathbf{L}^{-T}\boldsymbol{\psi}_i\big\|_2^2,
    \qquad
    \mathbf{G} = \mathbf{L}\mathbf{L}^{\top},
    \label{eq:christoffel_def}
\end{equation}
where $\boldsymbol{\psi}_i^{\top}$ denotes the $i$th row of $\boldsymbol{\Psi}$.
The stability of the least-squares solution is guaranteed when
\begin{equation}
    \frac{M}{\kappa \log M} = \mathcal{O}(1),
    \qquad
    \kappa = \max_{i} K_i,
    \label{eq:cohen_stability}
\end{equation}
which ensures that the Gram matrix~\eqref{eq:gram_matrix} is well-conditioned
with high probability.

\vspace{0.2em}
To reduce the influence of high-leverage samples, 
Liu \textit{et al.}~\cite{guo2019data} and Cohen 
proposed weighting each sample inversely to its Christoffel value,
leading to the \emph{Christoffel-weighted least-squares } formulation:
\begin{equation}
    \min_{\boldsymbol{c}}
    \; J_{\mathrm{CLS}}(\boldsymbol{c})
    = \frac{1}{M}
      \sum_{i=1}^{M}
      w_i\!\left(
      Y_i - \sum_{j=1}^{N} c_j\,\psi_j(\mathbf{x}_i)
      \right)^{\!2},
    \label{eq:cls_objective}
\end{equation}
with weights defined as
\begin{equation}
    w_i
    = \frac{M\,K_i^{-1}}{\sum_{j=1}^{M} K_j^{-1}}.
    \label{eq:cls_weights}
\end{equation}
The associated weighted normal equations are then given by
\begin{equation}
    \mathbf{G}_w\,\boldsymbol{c}
    = \frac{1}{M}\boldsymbol{\Psi}^{\top}\mathbf{W}\mathbf{Y},
    \qquad
    \mathbf{G}_w
    = \frac{1}{M}\boldsymbol{\Psi}^{\top}\mathbf{W}\boldsymbol{\Psi},
    \label{eq:cls_normal}
\end{equation}
where $\mathbf{W}=\mathrm{diag}(w_1,\ldots,w_M)$.
This weighting effectively regularizes $\mathbf{G}_w$ toward the identity
and reduces the coherence $\kappa$, improving the stability metric
$M/(\kappa\log M)$ introduced in~\eqref{eq:cohen_stability}.
{In the following analysis, this weighting formulation is used to examine how emphasizing low-leverage samples impacts the conditioning of the regression system.}

\subsection{Proposed Tempered Christoffel-Weighted Regression (T-CWLS)}
\label{sec:tcwls_theory}

\paragraph{Motivation.}
Although the Christoffel-weighted least-squares (CLS) formulation
in~\eqref{eq:cls_objective}--\eqref{eq:cls_normal}
stabilizes the regression system,
its fully inverse weighting may over-attenuate high-response samples,
leading to underestimation of the response tails.
In power-system applications,
accurate reconstruction of the tail region of the output probability distribution
(e.g., the 95th--99th percentiles) is critical,
as it governs rare-event probabilities and system risk.
To balance stability and tail fidelity,
we introduce a \emph{tempered} exponent~$\alpha$
that  controls the degree of Christoffel weighting.

\paragraph{Tempered weighting formulation.}
The proposed regression minimizes
\begin{equation}
    \min_{\boldsymbol{c}}
    \; J_{\mathrm{T\text{-}CWLS}}(\boldsymbol{c})
    = \frac{1}{M}
      \sum_{i=1}^{M}
      w_i(\alpha)\!
      \left(
        Y_i - \sum_{j=1}^{N} c_j\,\psi_j(\mathbf{x}_i)
      \right)^{\!2},
    \label{eq:tcwls_objective}
\end{equation}
where the tempered weights are defined as
\begin{equation}
    w_i(\alpha)
    = \frac{M\,K_i^{\alpha}}{\sum_{j=1}^{M} K_j^{\alpha}},
    \qquad
    K_i = \big\|\mathbf{L}^{-T}\boldsymbol{\psi}_i\big\|_2^2,
    \qquad
    \alpha\in\mathbb{R}.
    \label{eq:tcwls_weights}
\end{equation}
When $\alpha=0$, all samples are equally weighted
and~\eqref{eq:tcwls_objective} reduces to the ordinary least-squares (OLS) problem
of~\eqref{eq:ols_ddpce};
when $\alpha=-1$, it coincides with the inverse Christoffel weighting
of~\eqref{eq:cls_weights}.
The tempered values ($\alpha$)
smoothly trade off conditioning improvement and tail accuracy.
The weighted normal equations are
\begin{equation}
    \mathbf{G}_{\!w}(\alpha)\,\boldsymbol{c}
    = \frac{1}{M}\,\boldsymbol{\Psi}^{\top}\mathbf{W}(\alpha)\,\mathbf{Y},
    \qquad
    \mathbf{G}_{\!w}(\alpha)
    = \frac{1}{M}\,\boldsymbol{\Psi}^{\top}\mathbf{W}(\alpha)\boldsymbol{\Psi},
    \label{eq:tcwls_normal}
\end{equation}
where $\mathbf{W}(\alpha)=\mathrm{diag}(w_1(\alpha),\ldots,w_M(\alpha))$.

{In practice, the implementation involves computing the Christoffel values $K_i$, applying the tempered weighting $w_i(\alpha)$ for several candidate $\alpha$ values, solving the corresponding weighted regressions, and comparing the resulting surrogate outputs with the MCS reference to assess tail agreement. The analysis investigates how varying $\alpha$ influences numerical conditioning and tail accuracy, providing a balanced perspective between the unweighted OLS and the fully weighted CLS formulations.}

\subsection{Modeling Assumptions and Experimental Design}
{The study investigates how different regression weighting strategies within the DDSPCE framework affect the prediction accuracy of resilience-related metrics. The stochastic input vector is defined as
\begin{equation}
    \mathbf{x} = [P_{\mathrm{load}},\, T_{\mathrm{start}},\, T_{\mathrm{dur}}]^{\top},
\end{equation}
}
where $P_{\mathrm{load}}$ denotes the 24-hour load variation across the system, $T_{\mathrm{start}}$ represents the random starting time of the emergency event, and $T_{\mathrm{dur}}$ is its uncertain duration. For each realization $\mathbf{x}_i$, a deterministic scheduling problem is solved to obtain the corresponding system response
\begin{equation}
    Y_i = f(\mathbf{x}_i),
\end{equation}
where $Y_i$ is the total operational cost and unserved-load penalty reflecting the system’s resilience performance. The training dataset
\begin{equation}
    \mathcal{D} = \{(\mathbf{x}_i, Y_i)\}_{i=1}^{M}
\end{equation}
{is used to train and compare the OLS, CLS, and T-CWLS surrogates.}

{The analysis focuses on how the choice of regression weighting influences prediction accuracy and stability, rather than on extending the DDSPCE formulation itself.}

\color{black}
\section{Case study}
\subsection{Simulation Setup}

The test system consists of a distribution network connected to three MGs, each equipped with distributed energy resources and local loads. Under normal operating conditions, the MGs  can exchange power with the upstream grid. When an extreme event occurs, the connection to the main grid is interrupted, and the distribution system together with the MGs transitions into an emergency mode. During this emergency condition, the MGs must coordinate to supply the critical loads within the distribution system.
 
\begin{figure}[!b]
    \centering
    \includegraphics[width=0.9\linewidth]{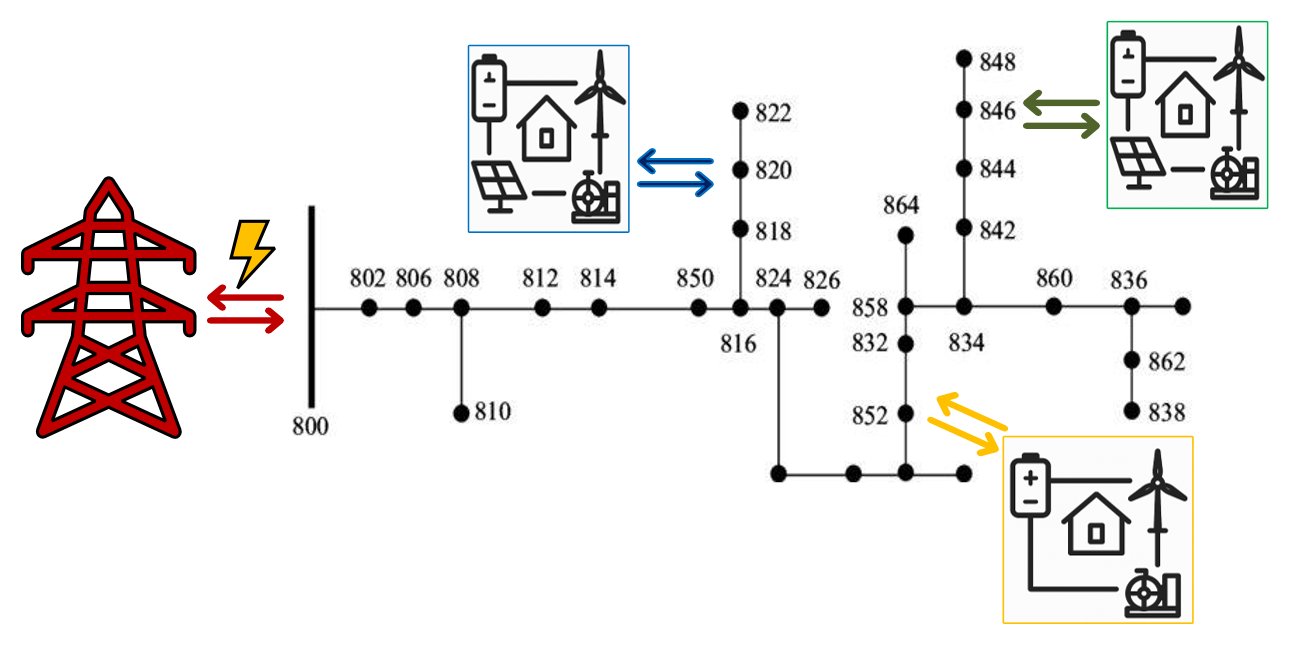}
    \caption{Modified IEEE~34-bus distribution system integrated with three MGs}
    \label{fig:test_system}
\end{figure}

A schematic of the modified IEEE~34-bus distribution system with three interconnected MGs is illustrated in Fig.~\ref{fig:test_system}.

\subsection{Result}
\begin{figure}[!t]
    \centering
    \subfloat[ ]{%
        \includegraphics[width=0.505\columnwidth,trim=40pt 0pt 10pt 0pt,clip]{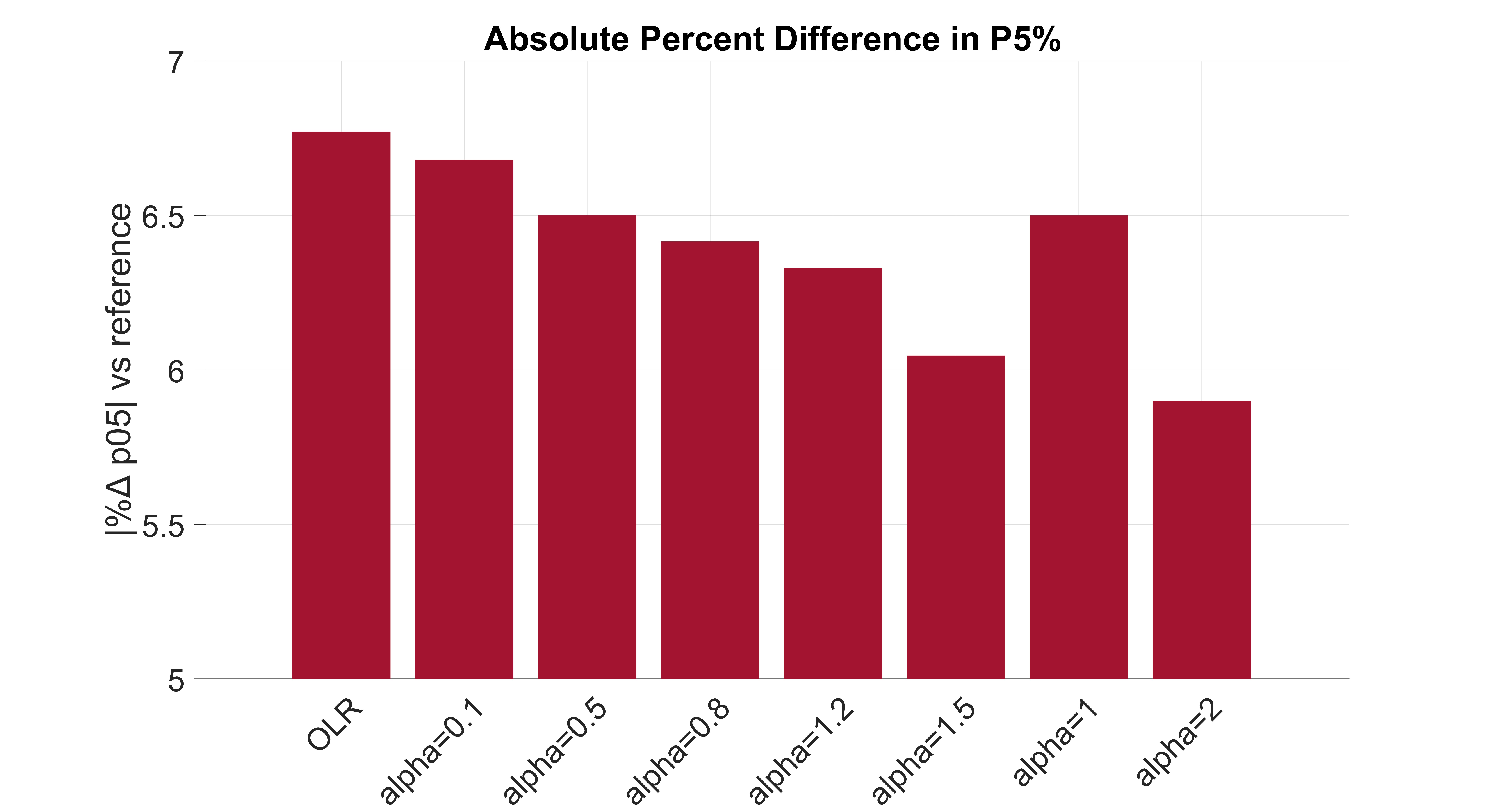}
        \label{fig:P5}}
        \hspace{-1cm} 
    \subfloat[ ]{%
        \includegraphics[width=0.505\columnwidth,trim=10pt 0pt 40pt 0pt,clip]{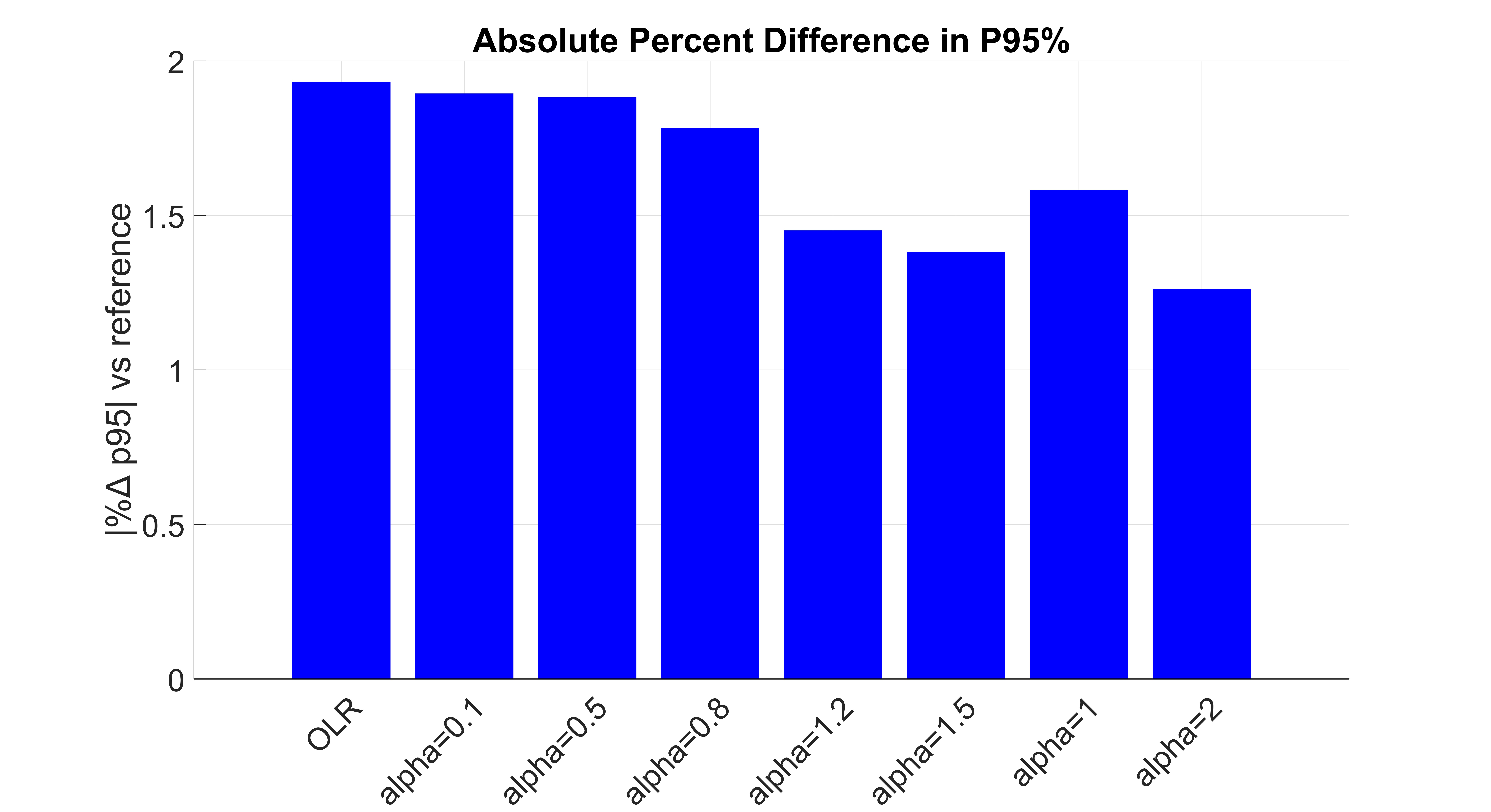}%
        \label{fig:P95}}
    \caption{Comparison of absolute percent deviations in the (a) 5th and (b) 95th percentiles of the surrogate-predicted output distributions for different weighting configurations.}
    \label{fig:abs_percent_deviation_combined}
\end{figure}

\begin{figure}[!t]
    \centering
    \subfloat[ ]{%
        
    \includegraphics[width=0.505\columnwidth,trim=40pt 0pt 10pt 0pt,clip]{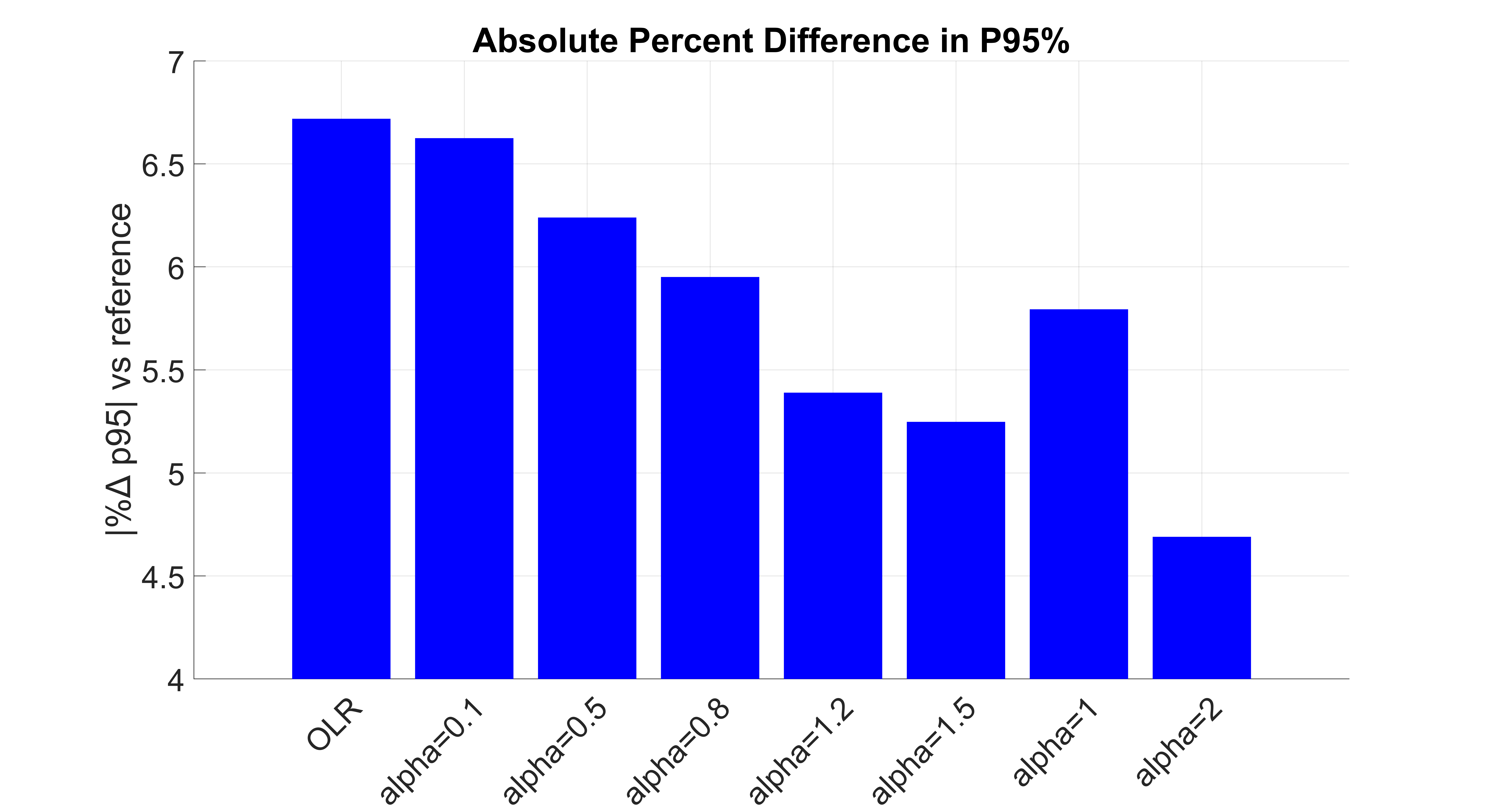}
        \label{fig:P95_OF}}
        \hspace{-1cm} 
    \subfloat[ ]{%
        \includegraphics[width=0.505\columnwidth,trim=10pt 0pt 40pt 0pt,clip]{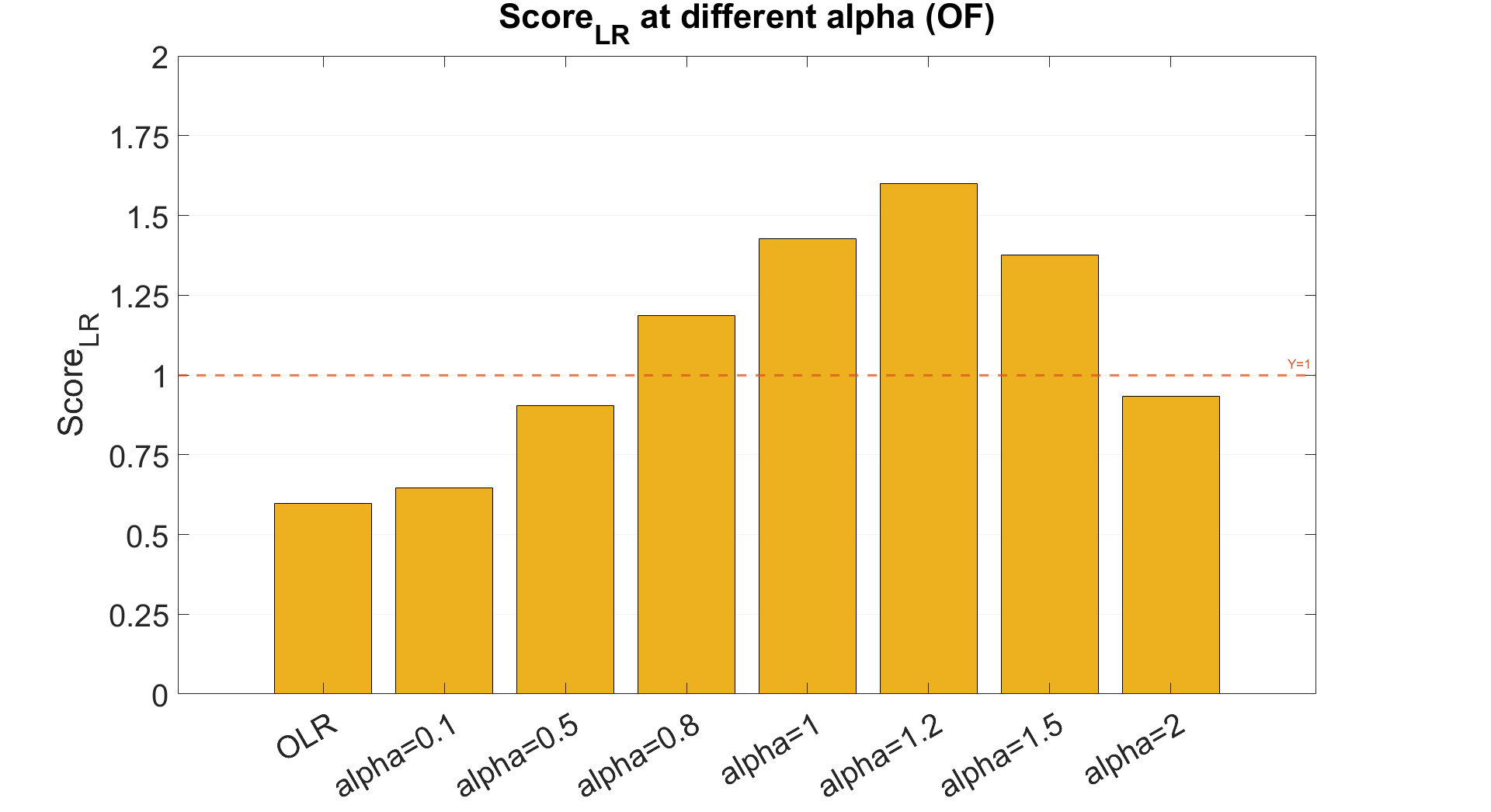}%
        \label{fig:Score_OF}}
    \caption{(a)Absolute percent deviation in the 95th percentile (\textit{P95\%}) 
for different~$\alpha$ values, (b)Variation of $\mathrm{Score}_{LR}$ with tempering exponent $\alpha$.}
    \label{fig:Fig3}
\end{figure}

Table~\ref{tab:wlars_results} reports how regression strategies affect the DDSPCE surrogate’s ability to reproduce the true model’s statistics. Deviations are measured relative to the MCS output; smaller values in the 5\% and 95\% columns indicate better accuracy in the lower and upper tails, which are central to resilience assessment.
The OLR,( $\alpha=0$), represents the unweighted formulation of DDSPCE. It yields unbiased estimates around the mean but underestimates the 5\% and 95\% percentiles, indicating limited accuracy in extreme responses. The low stability score $\mathrm{Score}_{\mathrm{LR}}$ further shows weak conditioning, which makes the estimated coefficients sensitive to training noise.
Applying the classical Christoffel weighting ($\alpha=1$) improves conditioning by reducing the influence of high-leverage samples. Although this enhances stability, it also weakens the contribution of tail observations, making the surrogate less responsive to rare events and leaving residual errors in the extreme quantiles.
The proposed tempered Christoffel weighting introduces a  control
parameter~$\alpha$ that allows the regression to transition smoothly between the
two limiting behaviors of OLR and fully weighted Christoffel regression.
This tempered weighting preserves sufficient emphasis on tail samples to recover
their statistical contribution, while still improving the conditioning of the
Gram matrix, as reflected by $\mathrm{Score}_{\mathrm{LR}}>1$.In fact, the stability index $\mathrm{Score}_{\mathrm{LR}}$ increases from 0.59 to values above~1.4, an improvement of more than 130\%. Figures~\ref{fig:P5} and~\ref{fig:P95} present the absolute percentage deviations in the estimated 5th and 95th percentiles of the output distribution with respect to the reference case(MC). These two metrics quantify how accurately each regression weighting configuration reproduces the lower and upper tails of the response.Specifically, the average 95th--percentile deviation is reduced by approximately 16\% and the 5th--percentile deviation by about 6\% compared with OLR.
These quantitative results confirm that tempering the Christoffel function
substantially enhances both regression robustness and the accuracy of tail
predictions.
Therefore, the proposed T--CWLS formulation provides a numerically stable and
tail--sensitive surrogate that reproduces the extreme behavior of the true MCS
model more faithfully than either the unweighted OLR or the conventional
Christoffel--weighted approach.
\begin{table}[!bt]
\centering
\caption{Statistical results for various $\alpha$ values for the distribution system's load shedding}
\label{tab:wlars_results}
\renewcommand{\arraystretch}{1.15}
\setlength{\tabcolsep}{5pt}
\begin{tabular}{lcccccc}
\toprule
\textbf{Case} & \textbf{5\% } & \textbf{95\% } & $\boldsymbol{\mu}$ & $\boldsymbol{\sigma}$ & \textbf{Score\_LR}\\
\midrule
OLR              & $-6.77\%$ & $1.93\%$  & $0.59\%$ & $1.03\%$  & $ 0.59$ \\
$\alpha:{0.1}$  & $6.73\%$  & $1.93\%$  & $0.6\%$  & $0.97\%$  & $0.646$  \\
$\alpha:{0.5}$  & $-6.5\%$  & $1.88\%$  & $0.6\%$  & $0.64\%$  & $0.905$  \\
$\alpha:{0.8}$  & $-6.42\%$ & $1.78\%$  & $0.65\%$ & $0.36\%$  & $1.187$  \\
$\alpha:{1.0}$  & $-6.5\%$  & $1.58\%$  & $0.69\%$ & $0.24\%$  & $1.47$   \\
$\alpha:{1.2}$  & $-6.33\%$ & $1.45\%$  & $0.74\%$ & $0.12\%$  & $1.607$  \\
$\alpha:{1.5}$  & $-6.05\%$ & $1.38\%$  & $0.73\%$ & $0.54\%$  &  $1.376$ \\
$\alpha:{2.0}$  &  $-5.9\%$ & $1.26\%$  & $0.83\%$ & $0.85\%$  & $0.993$   \\
\bottomrule
\end{tabular}
\end{table}
Furthermore, the influence of the tempering parameter~$\alpha$ on the objective function of the MGs' operation is illustrated in 
Figure~\ref{fig:Fig3}.
Figure~\ref{fig:P95_OF} shows that the absolute deviation in the 95th percentile decreases  as~$\alpha$ increases, reaching its minimum near~$\alpha=2$, which indicates that stronger tempering enhances the surrogate’s ability to capture the upper-tail behavior of the distribution. 
As illustrated in Figure~\ref{fig:Score_OF}, the minimum acceptable stability threshold, shown by the yellow reference line, marks the lower bound for reliable regression performance. The proposed tempered Christoffel approach maintains $\mathrm{Score}_{\mathrm{LR}}$ values above this threshold in some ~$\alpha$, whereas the unweighted OLR case falls below it. 
This demonstrates that the proposed weighting not only yields smaller tail deviations but also maintains stable behavior across~$\alpha$ values.

\color{black}
\section{Conclusion}
\label{sec:conclusion}
{This paper studied the impact of regression weighting intensity on the stability and tail prediction of DD–SPCE surrogates for resilience assessment. A tempered Christoffel weighted least squares formulation was examined, where the tempering exponent acts as a hyperparameter controlling the weighting applied during regression.} 
Case studies{ confirm that controlling the weighting intensity directly influences both numerical robustness and tail-prediction accuracy, supporting more reliable surrogate-based resilience assessment.}
{Future work will  explore practical approaches such as bootstrap-based stability analysis that resamples the available data to identify suitable values of the tempering exponent~$\alpha$. }

\bibliographystyle{IEEEtran}
\bibliography{Reference}
\end{document}